\documentclass[11pt,A4]{article}
\usepackage{graphics}
\begin{document}

\begin{center}
{\large \bf A General Approach to the Investigation of Soft Materials \\
by Thermodynamics Formalism}\\

\vspace{0.5cm}
{\large G. Oylumluoglu$^{*1,2}$, Fevzi B\"{u}y\"{u}kk\i l\i \c{c}$^{**1}$,  Dogan Demirhan}$^{***1}$\\
$^{1}$Ege University, Faculty of Science, Department of
Physics, Izmir-TURKEY.\\
$^{2}$Mugla University, Faculty of Arts and
Sciences, Department of Physics, Mugla-TURKEY.\\

\end{center}


\begin{abstract}
In this study, internal energy (U), electric field (E) and
particle number (N) which specify the system quantities i.e.
thermodynamical quantities for the proteins. In the frame of
thermodynamical formalism, the relation between the heat capacity
at effective field $C_E$ and the heat capacity at total dipole
moment $C_M$ and the relations for the increment of enthalpy
$\Delta H$, entropy $\Delta S$ and Gibbs energy $\Delta G$ which
come out in the dissolving of the proteins in water have been
obtained. By thinking about the present system being in the heat
bath the canonical distribution function, by considering the
system in heat and electric field bath the macro canonical
distribution function and once more by taking the system in heat
and particle number bath the macro canonical distribution function
have been calculated. Partition functions have been related to the
macro canonical quantities with the help of the free energy.
Understanding the structure of the proteins have been endeavoured
by fitting the theoretical calculations to the curves of
experimental results.

\vspace{0.5 cm} {\small Keywords: thermodynamics, classical
statistical mechanics, classical ensemble theory}

{\small PACS 05.70-a, 05.20-y, 05.20.Gg}

\vspace{.5cm} {\small $^*$ Corresponding
Authors:e-mail:oylum@sci.ege.edu.tr, Phone:+90 232-3881892
(ext.2363)
Fax:+90 232-3881892\\
\\
} \vspace{.5cm} {\small $^{**}$ e-mail:fevzi@sci.ege.edu.tr,
Phone:+90 232-3881892 (ext.2846)
Fax:+90 232-3881892\\
\\
} \vspace{.5cm} {\small $^{***}$ e-mail:dogan@sci.ege.edu.tr,
Phone:+90 232-3881892 (ext.2381)
Fax:+90 232-3881892\\
\\
}
\end{abstract}

\section{Introduction}

Now a days, a progress in the investigation of the soft materials
by the methods of statistical physics is recognized. It is not
possible to think a life without the soft materials; all
biological structures, molecules of the genetic code, proteins and
membranes contain soft materials. The soft materials form the
fundamental of life as well as being an important element of
future technological developments.

Statistical mechanical methods are the most practical methods in
obtaining the difference between the additional heat capacities
which aries in the dissolution of proteins (one of the soft
materials) in water and also for the calculation of
thermodynamical functions. In favour  of statistical
thermodynamics it is possible to explain the microscopic
quantities by relating them to the macroscopic experimental
results.

While thermodynamical variables of fluids are represented by the
physical quantities U (energy), V (volume) and N (particle
number), passing to the measurement results in the correspondence
to T (temperature), P (pressure) and $\mu$ (chemical potential)
consecutively \cite{Stanley}. In a similar manner, for the
magnetic systems the thermodynamics variables are represented by
the physical quantities U (energy), H (magnetic field) and N
(particle number) passing to the measurement leads to a
consecutive correspondence with T (temperature), M (magnetization)
and $\mu$ (chemical potential) \cite{Anderson}.

For thermodynamical quantities of the proteins; internal energy
(U), electric field (E) and particle number (N) could be taken as
macroscopic variables. When the system tends to thermodynamics
equilibrium, to relate with thermodynamics the system is
determined by internal energy parameter; temperature (T), electric
field parameter; dipole moment (M) and particle number parameter;
chemical potential ($\mu$).

\section{Relation between the heat capacities of proteins}

By taking electric field $E$ and total dipole moment $M$ as
thermodynamics variables the first law of thermodynamics for the
proteins could be written in the form:

\begin{equation}
  dU=TdS-MdE.
\label{Eq.2.1}
\end{equation}

Here, the electric field is not an external field but is used to
model the ice-like behavior of the water molecules around the
non-polar surface. The electric field is a result of the effective
behavior of the non-polar dissolvent that is applied to protein
unfolding. By taking the proteins as a soft material,
thermodynamical relation between the heat capacity $C_E$ at
effective electric field and heat capacity $C_M$ at constant total
dipole moment has been put forward. Rearranging \ref{Eq.2.1} one
gets:

\begin{equation}
  dQ=dU+MdE.
\label{Eq.2.2}
\end{equation}

To constitute such a relation if firstly the internal energy U is
written as the total differentials of the variable T and E one
could write

\begin{equation}
  U=U(T,E)\Longrightarrow dU(T,E)=(\frac{\partial U}{\partial
  T})_E dT+ (\frac{\partial U}{\partial
  E})_T dE
\label{Eq.2.3}
\end{equation}
and then substituting in \ref{Eq.2.1} gives

\begin{equation}
  dQ=(\frac{\partial U}{\partial T})_E dT+(\frac{\partial U}{\partial E})_T dE+
  MdE.
\label{Eq.2.4}
\end{equation}
Rewriting

\begin{equation}
  dQ=(\frac{\partial U}{\partial T})_E dT+[M+(\frac{\partial U}{\partial E})_T]dE.
\label{Eq.2.5}
\end{equation}
When heat capacities are defined; since $E=const.$ and $dE=0$,
\ref{Eq.2.5} becomes

\begin{equation}
  dQ=(\frac{\partial U}{\partial T})_E dT=C_E dT.
\label{Eq.2.6}
\end{equation}
Similarly when $M=const.$ and $dM=0$ \ref{Eq.2.5} then takes the
form

\begin{equation}
  dQ=(\frac{\partial U}{\partial T})_E dT+[M+(\frac{\partial U}{\partial E})_T]dE=C_MdT.
\label{Eq.2.7}
\end{equation}
After substituting the definitions of heat capacities given by
\ref{Eq.2.6} and \ref{Eq.2.7} into \ref{Eq.2.5} one obtains

\begin{equation}
 C_MdT=C_E dT+[M+(\frac{\partial U}{\partial E})_T]dE.
\label{Eq.2.8}
\end{equation}
Dividing both sides of \ref{Eq.2.8} by $dT$ for $M=const.$ and
since $\frac{dE}{dT}=(\frac{\partial E}{\partial T})_M$ one gets

\begin{equation}
 C_MdT=C_E dT+[M+(\frac{\partial U}{\partial E})_T](\frac{\partial E}{\partial T})_M.
\label{Eq.2.9}
\end{equation}
In order to express the term in the right hand side of this
equation in terms of experimentally measurable quantities,
\ref{Eq.2.1} could be written in the form:

\begin{equation}
 dS=\frac{1}{T} dU+\frac{M}{T} dE.
\label{Eq.2.10}
\end{equation}
When the total differential of the internal energy $U(T,E)$ is
taken as given in \ref{Eq.2.3} and substituted in \ref{Eq.2.10}

\begin{equation}
 dS=\frac{1}{T} [(\frac{\partial U}{\partial T})_E dT+(\frac{\partial U}{\partial E})_T dE] +\frac{M}{T} dE
\label{Eq.2.11}
\end{equation}
is obtained. Rearranging this equation

\begin{equation}
 dS=\frac{1}{T}(\frac{\partial U}{\partial T})_E dT +\frac{1}{T}[(\frac{\partial U}{\partial E})_T dE +\frac{M}{T}] dE.
\label{Eq.2.12}
\end{equation}
Similarly the total differential of the entropy $S(T,E)$ is
written in the form

\begin{equation}
  S=S(T,E)\Longrightarrow dS(T,E)=(\frac{\partial S}{\partial
  T})_E dT+ (\frac{\partial S}{\partial
  E})_T dE.
\label{Eq.2.13}
\end{equation}
Since corresponding terms in \ref{Eq.2.12} and \ref{Eq.2.13} must
be equal to each other:

\begin{equation}
  (\frac{\partial S}{\partial
  T})_E= \frac{1}{T} (\frac{\partial U}{\partial
  T})_E  and (\frac{\partial S}{\partial
  E})_T= \frac{1}{T} [(\frac{\partial U}{\partial
  E})_T+ M].
\label{Eq.2.14}
\end{equation}
Writing down the property of equality of the second derivatives
$[\frac{\partial}{\partial E} (\frac{\partial S}{\partial
T})_E=\frac{\partial}{\partial T} (\frac{\partial S}{\partial
E})_T]$ is substituted, \ref{Eq.2.14} becomes:

\begin{equation}
 \frac{\partial }{\partial E}[\frac{1}{T}(\frac{\partial
 U}{\partial T})_E]=\frac{\partial }{\partial T}\{\frac{1}{T} [(\frac{\partial U}{\partial E})_T +M] .
 \label{Eq.2.15}
\end{equation}
Taking the derivatives in \ref{Eq.2.15} and after simplification
one obtains:

\begin{equation}
 \frac{1}{T} \frac{\partial }{\partial E} (\frac{\partial U}{\partial T})_E=-\frac{1}{T^2} (\frac{\partial U}{\partial E})_T
 -\frac{M}{T^2}+ \frac{1}{T} \frac{\partial }{\partial T}(\frac{\partial U}{\partial E})_T+\frac{1}{T} (\frac{\partial M}{\partial T})_E .
 \label{Eq.2.16}
\end{equation}
Since this expression is a total differential it is independent of
the order of the derivative. Rewriting \ref{Eq.2.16}

\begin{equation}
 \frac{\partial }{\partial E} (\frac{\partial U}{\partial T})_E=-\frac{1}{T} (\frac{\partial U}{\partial E})_T
 -\frac{M}{T}+ \frac{\partial }{\partial T}(\frac{\partial U}{\partial E})_T+ (\frac{\partial M}{\partial T})_E
 \label{Eq.2.17}
\end{equation}
and then rearranging

\begin{equation}
 M+ (\frac{\partial U}{\partial E})_T= T (\frac{\partial M}{\partial T})_E
 \label{Eq.2.18}
\end{equation}
is obtained. Now substituting \ref{Eq.2.18} in \ref{Eq.2.9} one
finds:

\begin{equation}
 C_M=C_E+ T (\frac{\partial M}{\partial T})_E (\frac{\partial E}{\partial
 T})_M.
 \label{Eq.2.19}
\end{equation}
Using the change rule of the total differential $(\frac{\partial
M}{\partial T})_E (\frac{\partial T}{\partial E}) _M
(\frac{\partial E}{\partial M}) _T=-1$ one could write
$(\frac{\partial M}{\partial T})_E =(\frac{\partial E}{\partial
T}) _M (\frac{\partial M}{\partial E}) _T$. Substituting this
expression in \ref{Eq.2.19} gives:

\begin{equation}
 C_M=C_E- T (\frac{\partial M}{\partial E})_T [(\frac{\partial E}{\partial
 T})_M]^2.
 \label{Eq.2.20}
\end{equation}
By taking use of Maxwell relations written for soft materials
(proteins) and with the purpose of expressing in terms of the
measurable parameters from thermodynamics the following definition
is given

\begin{equation}
 -(\frac{\partial S}{\partial M})_T=(\frac{\partial E}{\partial
 T})_M=\alpha_M
 \label{Eq.2.21}
\end{equation}
which transforms \ref{Eq.2.19} into:

\begin{equation}
 C_M=C_E-T \alpha_M^2 (\frac{\partial M}{\partial E})_T.
 \label{Eq.2.22}
\end{equation}
Similarly when the definition

\begin{equation}
 -(\frac{\partial M}{\partial E})_T=\frac{1}{\gamma_T}
 \label{Eq.2.23}
\end{equation}
is written down \ref{Eq.2.23} now takes the form:

\begin{equation}
 C_M=C_E-T \frac{\alpha_M^2}{\gamma_T}.
 \label{Eq.2.24}
\end{equation}

\section{Thermodynamics of enthalpy, entropy and gibbs energy increments}
The first law of thermodynamics for the soft materials has been
written by \ref{Eq.2.1}. When S and M are taken as direct
variables, change in enthalpy reads

\begin{equation}
 dH=T dS+E dM.
 \label{Eq.3.1}
\end{equation}
The enthalpy increment, after taking the total differentials of
enthalpy $H=H(S,M)$ and entropy $S=(T,M)$, is written in the form:

\begin{equation}
dH=(\frac{\partial H}{\partial S})_MdS+(\frac{\partial H}{\partial
M})_S dM
 \label{Eq.3.2}
\end{equation}

\begin{equation}
 dS=(\frac{\partial S}{\partial T})_M dS+(\frac{\partial
S}{\partial M})_S dM. \label{Eq.3.3}
\end{equation}
By substituting \ref{Eq.3.2} and \ref{Eq.3.3} in \ref{Eq.3.1}

\begin{equation}
 dH=(\frac{\partial H}{\partial S})_M (\frac{\partial S}{\partial T})_M dT+
 [(\frac{\partial H}{\partial S})_M (\frac{\partial S}{\partial M})_T+(\frac{\partial H}{\partial M})_S] dM. \label{Eq.3.4}
\end{equation}
is obtained. The expressions $T=(\frac{\partial H}{\partial
S})_M$, $E=(\frac{\partial H}{\partial M})_S$ and
$\frac{C_M}{T}=(\frac{\partial S}{\partial T})_M$ are substituted
in \ref{Eq.3.4} and than integration of the equation leads to the
enthalpy increment:

\begin{equation}
 \Delta H=\int_{T_t}^T C_M dT+\int_{M(T_t)}^{M_(T)}[E+T(\frac{\partial S}{\partial M})_T] dM. \label{Eq.3.5}
\end{equation}

For the entropy increment; the total differential of entropy
$S=S(T,M)$ is taken and $(\frac{\partial E}{\partial
T})_M=-(\frac{\partial S}{\partial M})_T$ and
$\frac{C_M}{T}=(\frac{\partial S}{\partial T})_M$ are substituted,
which after integration gives:

\begin{equation}
 \Delta S=\int_{T_t}^T \frac{C_M}{T} dT+\int_{M(T_t)}^{M_(T)}(\frac{\partial S}{\partial M})_T dM. \label{Eq.3.6}
\end{equation}

By taking use of the first law of thermodynamics given in
\ref{Eq.2.1} which has been written for the soft materials and
taking into consideration the parameters U, E and N of the soft
materials, change in the Gibbs energy could be obtained as

\begin{equation}
 dG=-SdT+EdM+\mu dN.
\label{Eq.3.7}
\end{equation}
With a similar approach, the Gibbs energy increment is determined
in the form

\begin{equation}
 \Delta G=-\int_{T_t}^T S dT+\int_{M(T_t)}^{M_(T)}E dM+\int_{N(T_t)}^{N_(T)} \mu dN \label{Eq.3.8}
\end{equation}
by taking the total differentials of Gibbs energy $G=G(T,M)$ and
total dipole moment $M=M(T,E)$. These expressions together with
$E=(\frac{\partial G}{\partial M})_T$ and $-S=(\frac{\partial
G}{\partial T})_M$ are substituted in \ref{Eq.3.7} and then the
equation is integrated.

\section{Establishment of the relation between thermodynamics and statistical mechanics}
Proteins is a common name of complex macromolecules which are
formed by a collection of a great number of amino-acids and which
play vital roles in the life times of all of the living creatures.
As a physical object, the protein molecule is a rather big polymer
which is formed by thousands of atoms; in other words it is a soft
matter such that from the physical point of view this is a
macroscopic system.

The three dimensional structure of the proteins depends on
different environmental factors inside the cell. Unfolding
proteins return to their natural state when the environmental
factors are removed, in manner of exhibiting the fact that the
three dimensional structure of a protein solely depends on its
information of the amino-acid regularity. the probable
configurations of this three dimensional structure are in
astronomically large order. In the thermodynamical limit
($N\rightarrow \infty$, $E\rightarrow \infty$  and $\frac{N}{E}=
const.$) it is not possible to make calculations by taking into
consideration each particle in the macroscopic system formed from
the proteins, even by using the computer existing on Earth with
highest operation capacity. Thus combination of thermodynamics and
statistical physics is a necessity. In the investigation of these
type of system with the help of statistical physics, the existence
of a great number of systems that are identical to the system
under consideration is taken into account and using the
statistical methods for these system, the most probable state in
which the real system could enter is the task of determination.

Statistical ensemble mean approach is one of the appealed methods
in statistical mechanics. In order to constitute the statistical
ensembles in the system that are interacting with its surroundings
the system has to be taken into account together with its
surroundings. Distribution functions are determined in the
equilibrium state using the Lagrange multipliers method from
statistical ensemble theory.

\subsection{A general approach to the calculation of distribution function, entropy and thermodynamical potential}

\subsubsection{Distribution function and partition function}

For a general approach to the calculations, let the parameters U,
E, N of the thermodynamical system under consideration, fluctuate
together. In order words, the system could exchange energy,
particles as well as electric field.  Probability function P which
gives the distribution of the system in to microstate is a
function of n, E and N. Here n is a quantum number arising from
the energy dependence. Number of particles N and electric field E
also depend on energy. The entropy S of the system is given by:

\begin{equation}
 S=-k\sum_{n,E,N} P_{n,E,N} ln P_{n,E,N}.
\label{Eq.4.1.1}
\end{equation}
The following equations for the total probability, the average
energy $\overline{U}$, average electric field $\overline{E}$ and
average number of particles $\overline{N}$ of the system could
respectively be written;

\begin{equation}
 \sum_{n,E,N} P_{n,E,N}= 1
\label{Eq.4.1.2}
\end{equation}

\begin{equation}
 \sum_{n,E,N} P_{n,E,N} U_{n,E,N}= \overline{U}
\label{Eq.4.1.3}
\end{equation}

\begin{equation}
 \sum_{n,E,N} P_{n,E,N} E_n= \overline{E}
\label{Eq.4.1.4}
\end{equation}

\begin{equation}
 \sum_{n,E,N} P_{n,E,N} N_n= \overline{N}
\label{Eq.4.1.5}
\end{equation}

Using the maximum entropy principle, the distribution function
$P_{n,E,N}$ of the system could be determined by Lagrange
multipliers method. Thus the problem is; to make the entropy given
by \ref{Eq.4.1.1} maximum under the constraints given by the
\ref{Eq.4.1.2}, \ref{Eq.4.1.3}, \ref{Eq.4.1.4} and \ref{Eq.4.1.5}.
In the mathematical formalism of Lagrange multipliers methods, the
expression;

\begin{eqnarray}
  Q=-k \sum_{n,E,N} P_{n,E,N} ln P_{n,E,N}+ \alpha_1 (1- \sum_{n,E,N}
 P_{n,E,N})+
   \alpha_2 (\overline{U}- \sum_{n,E,N}
 P_{n,E,N} U_{n,E,N})+ \nonumber \\
  \alpha_3 (\overline{E}- \sum_{n,E,N}
 P_{n,E,N} E_n)+
   \alpha_4 (\overline{N}- \sum_{n,E,N}
 P_{n,E,N} N_n) \label{Eq:4.1.6}
\end{eqnarray}
should be maximum, where $\alpha_1$, $\alpha_2$, $\alpha_3$,
$\alpha_4$ are undetermined Lagrange multipliers. Differentiating
Q with respect to $P_{n,E,N}$ and equating to zero one obtains:

\begin{equation}
 P_{n,E,N}=exp\{-(\alpha+ \beta U_{n,E,N}+ \theta E_n+ \gamma N_n)\}
\label{Eq.4.1.7}
\end{equation}
where

\begin{equation}
 \alpha= \frac{k+ \alpha_1}{k},\beta=\frac{\alpha_2}{k},
 \theta=\frac{\alpha_3}{k}, \gamma=\frac{\alpha_4}{k}.
\label{Eq.4.1.8}
\end{equation}
Taking under consideration the condition in \ref{Eq.4.1.2};
\ref{Eq.4.1.7} leads to

\begin{equation}
 exp \alpha= Z(\beta, \theta, \gamma)
\label{Eq.4.1.9}
\end{equation}
where $Z(\beta, \theta, \gamma)$ is the partition function of the
system and is given by

\begin{equation}
 Z(\beta, \theta, \gamma)=\sum_{n,E,N} exp{-(\beta U_{n,E,N}+ \theta E_n+ \gamma
 N_n)}.
\label{Eq.4.1.10}
\end{equation}
Making use of \ref{Eq.4.1.9} the distribution function now becomes

\begin{equation}
 P_{n,E,N}=\sum_{n,E,N} \frac{exp{-(\beta U_{n,E,N}+ \theta E_n+ \gamma
 N_n)}}{Z(\beta, \theta, \gamma)}
\label{Eq.4.1.11}
\end{equation}
where $\beta, \theta, \gamma$ are functions of the parameters U, E
and N which describe the system. In the rest of the manuscript the
distribution function given by \ref{Eq.4.1.11} will be referred as
macro canonical generalized distribution function.

\subsubsection{Entropy}

When the distribution function given by \ref{Eq.4.1.11} is
substituted in \ref{Eq.4.1.1} and taking into account that macro
quantities U, E, N peak perfectly such that they could be taken
instead of their average values, the entropy of the system
becomes:

\begin{equation}
 S=k \beta U+ k \theta E+ k \gamma N+ k ln Z(\beta, \theta, \gamma)
\label{Eq.4.2.12}
\end{equation}

\subsubsection{Determination of the Lagrange multipliers}

With the purpose of obtaining the physical counterparts of the
multipliers $\beta$, $\theta$ and $\gamma$, the definition of
temperature $T$, total dipole moment $M$ and chemical potential
$\mu$ are made;

\begin{equation}
 \frac{1}{T}= (\frac{\partial S}{\partial U})_{E,N}
\label{Eq.4.3.13}
\end{equation}

\begin{equation}
 M= T (\frac{\partial S}{\partial E})_{U,N}
\label{Eq.4.3.14}
\end{equation}

\begin{equation}
 \mu= -T (\frac{\partial S}{\partial N})_{U,E}.
\label{Eq.4.3.15}
\end{equation}
Furthermore, after differentiation it is easy to show that;

\begin{equation}
 \frac{1}{Z}(\frac{\partial Z}{\partial \beta})_{\theta, \gamma}=
 -\overline{U}=-U
\label{Eq.4.3.16}
\end{equation}

\begin{equation}
 \frac{1}{Z}(\frac{\partial Z}{\partial \theta})_{\beta, \gamma}=
 -\overline{E}=-E
\label{Eq.4.3.17}
\end{equation}

\begin{equation}
 \frac{1}{Z}(\frac{\partial Z}{\partial \gamma})_{\beta, \gamma}=
 -\overline{N}=-N.
\label{Eq.4.3.18}
\end{equation}
In the above equations, instead of the average values U, E and N
are taken.

Substituting \ref{Eq.4.2.12} in \ref{Eq.4.3.13} and making use of
\ref{Eq.4.3.16} one ends up with:

\begin{equation}
 \beta=\frac{1}{k T}.
\label{Eq.4.3.19}
\end{equation}
Similarly substituting of \ref{Eq.4.2.12} in \ref{Eq.4.3.14} and
\ref{Eq.4.3.15} respectively leads to:

\begin{equation}
 \theta=\frac{M}{k T}
\label{Eq.4.3.20}
\end{equation}

\begin{equation}
 \gamma=-\frac{\mu}{k T}.
\label{Eq.4.3.21}
\end{equation}
In the content of these calculations, distribution function
\ref{Eq.4.1.11} takes form:

\begin{equation}
 P_{n,E,N}= \frac{exp{-\beta (U_{n,E,N}+ M E_n- \mu
 N_n)}}{Z(T, M, \mu)}
\label{Eq.4.3.22}
\end{equation}
and for the partition function one obtains:

\begin{equation}
 Z_{n,E,N}= \sum_{n,E,N} exp{-\beta (U_{n,E,N}+ M E_n- \mu
 N_n)}.
\label{Eq.4.3.23}
\end{equation}
On the other hand for the entropy given by \ref{Eq.4.2.12}

\begin{equation}
 S=\frac{U}{T}+ \frac{M E}{T}- \frac{\mu N}{T}+ k ln Z
\label{Eq.4.3.24}
\end{equation}
could be written.

\subsubsection{Thermodynamical potential}

Commencing with the entropy namely \ref{Eq.4.3.24} and considering

\begin{equation}
 \Phi(T,M,\mu)=-k T ln Z
\label{Eq.4.4.25}
\end{equation}
as thermodynamical potential one ends up with

\begin{equation}
 \Phi(T, M, \mu)= F+ M E- \mu N
\label{Eq.4.4.26}
\end{equation}
where $F=U-TS$ is the Helmholzt free energy. \ref{Eq.4.4.26} could
also be written in a different form:

\begin{equation}
 \Phi(T, M, \mu)= G- \mu N
\label{Eq.4.4.27}
\end{equation}
where $G= U- T S+ M E$ is the Gibbs free energy.

\subsection{Obtaining the canonical distributions from the generalized distribution}
The micro canonical, canonical and macro canonical distributions
that are frequently met in the literature \cite{Turner} can be
obtained as the special cases of the generalized distribution,
which is given by \ref{Eq.4.1.11}.

\subsubsection{System in a heat bath: canonical distribution}
Thermodynamical systems which obey micro canonical distributions
are systems isolated from energy, electric field and particle
number exchanges. Therefore, energy has a unique value.

In the thermodynamical systems governed by the canonical
distribution there is energy exchange but the system is isolated
from electric field and particle number exchanges. For this system
$\alpha_1 \neq 0$, $\alpha_2 \neq 0$, $\alpha_3= 0$, and
$\alpha_4= 0$. When the corresponding multipliers are substituted
in \ref{Eq.4.1.11} the distribution function takes the form;

\begin{equation}
 P_n= \frac{1}{Z} exp(-\beta U_n)
\label{Eq.4.2.28}
\end{equation}
and the partition function \ref{Eq.4.1.10} becomes

\begin{equation}
 Z=\sum_n exp-\beta U_n.
\label{Eq.4.2.29}
\end{equation}
One order hand the entropy expression given by \ref{Eq.4.2.12}
reduces to

\begin{equation}
S=\frac{U}{T}+ k ln Z.
 \label{Eq.4.2.30}
\end{equation}
Meanwhile the thermodynamical potential of the system could be
written as

\begin{equation}
 F=U- T S
\label{Eq.4.2.31}
\end{equation}
where $F=-k T ln Z$ is Helmholtz free energy.

Statistical mechanics plays the role of bridging the understanding
of the physical quantities in terms of their microscopic
counterparts. The energy of the proteins system is determined by:

\begin{equation}
 U=-\frac{1}{Z} \frac{\partial }{\partial \beta} Z.
\label{Eq.4.2.32}
\end{equation}
The heat capacity at effective electric field has been calculated
from the expressions of partition function and internal energy
giving;

\begin{equation}
 C_E=-k \beta^2(\frac{\partial U}{\partial \beta})_E.
\label{Eq.4.2.33}
\end{equation}
Heat capacity at constant total dipole moment is obtained by
making use of expression \cite{Reif}

\begin{equation}
 C_M=C_E-T (\frac{\partial M}{\partial E})_T [(\frac{\partial E}{\partial T})_M]^2.
\label{Eq.4.2.34}
\end{equation}

In a canonical ensemble, connection with thermodynamics is not
established by entropy but with Helmholtz free energy. The
following relations are written between Helmholtz free energy and
the thermodynamical quantities;

Entropy;\begin{equation}
  S=-(\frac{\partial F}{\partial T})_E
\label{Eq.4.2.35}
\end{equation}

Total dipole moment;\begin{equation}
 M=-(\frac{\partial F}{\partial E})_T.
\label{Eq.4.2.36}
\end{equation}

\subsubsection{System in a heat and electric field bath: macro canonical distribution}
Let us take into account the system where energy U and electric
field E fluctuate but the number of particles is constant. For
this case $\alpha_1 \neq 0$, $\alpha_2 \neq 0$, $\alpha_3 \neq 0$,
and $\alpha_4= 0$. In a similar manner to the above mentioned
distributions, the distribution function

\begin{equation}
 P_{n,E}= \frac{1}{\Xi} exp-\beta (U_{n,N}-M E_n)
\label{Eq.4.2.37}
\end{equation}
the partition function

\begin{equation}
 \Xi=\sum_{n,E} exp-\beta (U_{n,N}-M E_n)
\label{Eq.4.2.38}
\end{equation}
the entropy

\begin{equation}
S=\frac{U}{T}+ \frac{M E}{T}+ k ln \Xi
 \label{Eq.4.2.39}
\end{equation}
and thermodynamical potential

\begin{equation}
 \Phi=F+M E
\label{Eq.4.2.40}
\end{equation}
are obtained, where $\Phi=-k T ln \Xi$.

\subsubsection{System in a heat and particle bath: macro canonical distribution}
In thermodynamical systems obeying the macro canonical
distribution, the system is in a heat bath and material bath. For
such a system $\alpha_1 \neq 0$, $\alpha_2 \neq 0$, $\alpha_3= 0$,
and $\alpha_4\neq 0$. When the corresponding multipliers are
determined and substituting in \ref{Eq.4.1.11}, distribution
function takes the form:

\begin{equation}
 P_{n,N}= \frac{1}{\Xi} exp-\beta (U_{n,N}-\mu N_n).
\label{Eq.4.2.41}
\end{equation}
For the systems obeying macro canonical distribution,
\ref{Eq.4.1.10}, which gives the partition function, reads

\begin{equation}
 \Xi=\sum_{n,N} exp-\beta (U_{n,N}-\mu N_n)
\label{Eq.4.2.42}
\end{equation}
the entropy equation now becomes

\begin{equation}
S=\frac{U}{T}- \frac{\mu N}{T}+ k ln \Xi
 \label{Eq.4.2.43}
\end{equation}
and thermodynamical potential of the system is

\begin{equation}
 \Phi=F- \mu N.
\label{Eq.4.2.44}
\end{equation}

In the macro canonical ensemble, in the first approximation, when
the equation expressing the canonical ensemble is extended to
macro canonical case without putting any restriction on the
particle number the, following equation could be used
\cite{Greiner};

\begin{equation}
 \Phi(T, E, \mu)=\sum_{N=0}^ \infty\frac{1}{N!}[exp (\frac{\mu}{k T}) Z(T, E,
 N=1)]^N.
\label{Eq.4.2.45}
\end{equation}

As it is recognized, the partition function is summed over by
weighing with Gibbs factor which has been used in the calculation
of undistinguishable particles. When the sum in \ref{Eq.4.2.45} is
calculated;

\begin{equation}
 \Phi(T, E, z)= exp {z Z(T, E,N=1)}
 \label{Eq.4.2.46}
\end{equation}
is obtained. In this equation, $\mu$ being the chemical potential,
the term $z=exp(\beta \mu)$ corresponds to the pH of the system
and it is a weighing factor. The energy of the system is given by

\begin{equation}
 U=-(\frac{\partial ln \Xi}{\partial \beta})_{\mu, E}
 \label{Eq.4.2.47}
\end{equation}
and it is determined by differentiation after substitution of the
partition function. In terms of this grand potential, the
expressions for the other thermodynamical quantities are given by;

Entropy;\begin{equation}
  S=-(\frac{\partial \Phi}{\partial T})_{\mu, E}
\label{Eq.4.2.48}
\end{equation}

Total dipole moment;\begin{equation}
 M=-(\frac{\partial \Phi}{\partial E})_{\mu, T}
\label{Eq.4.2.49}
\end{equation}

Particle number;\begin{equation}
 N=-(\frac{\partial \Phi}{\partial \mu})_{T,E}.
\label{Eq.4.2.50}
\end{equation}

Enthalpy, Gibbs energy and entropy increments and physical
quantities $\epsilon$, $b$ and $\mu$ in \ref{Eq.4.2.33},
\ref{Eq.4.2.34}, \ref{Eq.4.2.35} and \ref{Eq.4.2.36} that has been
obtained in this semi phenomenological theory are determined by
fitting experimental results given by \cite{Privalov,Creighton}.

\section{Conclusions}\label{Sect.5}
In this study, for the soft matters in the frame of thermodynamics
formalism, the relation between the heat capacities in
\ref{Eq.2.24}, $\Delta H$ enthalpy, $\Delta S$ entropy and Gibbs
energy $\Delta G$ increments have been calculated and presented in
\ref{Eq.3.5}, \ref{Eq.3.6} and \ref{Eq.3.8} respectively.
Establishment of the relation between the thermodynamics and
statistical mechanics ha been done with the help of the partition
function and given by \ref{Eq.4.3.24}, \ref{Eq.4.2.30},
\ref{Eq.4.2.38} and \ref{Eq.4.2.43}. Calculation of the partition
functions for proteins (soft materials) with multi molecules
exhibits difficulties. Therefore the partition function of a
single molecules is obtained with the mean field approximation and
then multiplied by N. The partition function is related to the
parameters $\epsilon$, $b$ and $\mu$ which determined the
properties of the molecule. In \ref{Eq.2.24} presented here; $C_M$
is taken from experimental data, $C_E$ is theoretically calculated
and the above mentioned parameters are obtained by fitting to the
experimental results.

On the other hand in the calculation of the increments of enthalpy
$\Delta H$, entropy $\Delta S$ and Gibbs energy $\Delta G$, one
proceeds in a similar manner. As an example, in the calculation of
the entropy increments, as mentioned above $ C_M$ is taken from
the experimental data and the term related with entropy is
obtained from \ref{Eq.4.2.35}. As a conclusion one could say that;
the structure of the proteins (soft materials) might be understood
by the thermodynamics, statistical mechanics and experimental
studies triplet using this semi phenomenological approach.

The significance of the electric field is to model the ice-like
behavior exhibited by the water molecules around the non-polar
surfaces which is a kind of stiffness \cite{Bakk}. Here this is
not a real external electric field but the effective behavior of
the non-polar dissolute enforced upon in the protein unfolding. It
could also be possible to approach to this problem by considering
the protein as a rigid body. However, in the proteins which are
example of the soft matters, the benefit of determination of the
expression which displays the thermodynamical relation between the
heat capacity at effective electric field $C_E$ and the heat
capacity at constant total dipole moment $C_M$ is evident.

As it could be seen from \ref{Eq.2.24} which shows the
thermodynamical relation between the heat capacity at effective
electric field $C_E$ and the heat capacity at constant total
dipole moment $C_M$, the difference between the heat capacities
depends linearly on temperature. However results of experimental
studies indicate a nonlinear dependence. This in turn implies the
necessity of taking into account the internal structure of
proteins. $C_E$ given in \ref{Eq.2.24} could easily be calculated
theoretically and if $C_M$ expression is obtained experimentally
then this enables the system to be investigated in terms of
thermodynamical quantities.

\end{document}